\documentclass{optica-article}
\journal{opticajournal} 

\articletype{Research Article}
\usepackage{float}
\usepackage{changes}
\usepackage{lineno}
\usepackage{svg}

\begin{document}

\title{Conditional neural holography: \\a distance-adaptive CGH generator}

\author{Yuto Asano,\authormark{1} Kenta Yamamoto,\authormark{2} Tatsuki Fushimi,\authormark{3} and Yoichi Ochiai\authormark{4}}

\address{
\authormark{1,2} Graduate School of Comprehensive Human Sciences, University of Tsukuba, Tsukuba 305-8550 Ibaraki, Japan\\\authormark{3,4}R\&D Center for Digital Nature, University of Tsukuba, Tsukuba, 305-8550 Ibaraki, Japan\\\authormark{4}Pixie Dust Technologies, Inc., Chiyoda-ku, 101-0061 Tokyo, Japan}

\email{\authormark{4}wizard@slis.tsukuba.ac.jp}



\begin{abstract*} 
A convolutional neural network (CNN) is useful for overcoming the trade-off between generation speed and accuracy in the process of synthesizing computer-generated holograms (CGHs). However, methods using a CNN have limited applicability as they cannot specify the propagation distance when synthesizing a hologram. We developed a distance-adaptive CGH generator that can generate CGHs by specifying the target image and propagation distance, which comprises a zone plate encoder stage and an augmented HoloNet stage. Our model is comparable to that of prior CNN methods, with a fixed distance, in terms of performance and achieves the generation accuracy and speed necessary for practical use.

\end{abstract*}

\section{Introduction}
Computer-generated holography (CGH)\cite{cgh_review} is a promising technology for reconstructing fully three-dimensional images, with potential applications in the fields of augmented reality and virtual reality. Key applications of this technique include holographic near-eye displays\cite{MS_near-eye_holo, Kim2022, near-eye-aberration} and holographic projectors\cite{Makowski:12,Buckley:11,Shimobaba:13}. Compared with conventional near-eye displays, holographic near-eye displays achieve higher spatial resolution and make it easier to perceive depth cues. Holographic projectors could potentially be made considerably smaller than conventional ones because they do not require lenses. To utilize CGH in such displays, highly accurate holograms need to be generated in real-time. Additionally, they should adapt to various propagation distances (i.e., the distance between the hologram and reconstructed image) because they can vary depending on the wearer and/or circumstances. 

Since Gabor invented the concept of holography\cite{gabor}, a wide range of methods to generate CGH have been proposed. Initial efforts were focused on enhancing the accuracy of two-dimensional CGH using iterative calculations. For example, the Gerchberg and Saxton algorithm\cite{GS} is a classical approach, and gradient descent methods have been proposed by Zhang et al.~\cite{Zhang:17}, Chakravarthula et al.~\cite{chakravarthula2019wirtinger}, and Peng et al.~\cite{citl}. While  iterative methods can generate accurate CGH, there is a significant trade-off between the generation time and the accuracy of the reconstructed image. Generating CGH in real-time with iterative methods usually compromises reconstructed image accuracy. While single calculation methods like direct phase addition and correction (DPAC) \cite{MS_near-eye_holo} offer an alternative, they lack the precision of iterative techniques. Neither iterative nor single calculation methods achieve a PSNR above 30 dB or exceed 30 fps.

Considering these circumstances, methods employing a convolutional neural network (CNN) have begun to attract attention~\cite{shimobaba2022deep}. An early attempt to generate CGH using a CNN was demonstrated by Horisaki et al.~\cite{Horisaki:18}, where a simple U-Net~\cite{u_net} was used. Wang et al.~constructed a new CNN model called Y-net for CGH generation and successfully achieved improvements in accuracy\cite{Wang:19}. Peng et al.~proposed a parameterized network model called HoloNet\cite{citl}, which was further improved by Dong et al.~in 2023\cite{Dong:23}. Yu et al.~also introduced a CNN model for generating high quality 2D holograms~\cite{yu2022phase}. Shi et al.~developed CNN models for 3D CGH generation with a high accuracy~\cite{shi2021towards,shi2022end}. These reported CNN-based methods overcome the trade-off between CGH generation speed and accuracy.



However, a method for generating CGH that transcends this trade-off and accommodates various propagation distances has yet to be explored. Although CNN models have addressed the trade-off for a fixed propagation distance, they cannot generate CGH for multiple propagation distances in real-time, because they need to be re-trained whenever the propagation distance changes. Shui et al.~\cite{Shui:22} have demonstrated a CNN that can adapt to various distances within a range of 0.3 m to 0.32 m; with 20 discrete steps within the range; however, the variable range is limited and the number of steps is low. A range of adjustments on the order of 0.1 m is reasonably expected in holographic projectors, and there is a need to examine CNN models that adapt to such ranges with finer resolution. 

To address this issue, we propose a distance-adaptive CGH generator, which is a CNN model that can accept not only the desired target image but also the propagation distance as inputs, generating a CGH that reproduces a high-precision image when reconstructed at the given propagation distance.
We assessed the suitability of various CNN models for our study and discovered that some models could produce CGH with the same accuracy as conventional ones, even with propagation distances changing at over 100 points within a range of more than 0.1 m.

As detailed in Methods section, this distance-adaptive CGH generator comprises two stages. Although this design is not flawless and future research may develop single-stage models, none have been identified yet. Thus, this paper serves as an initial step in the research of distance-adaptive CGH generation.

\section{Methods}
\subsection{Principle of CGH Generation}
 CGH is based on the principle of light diffraction~\cite{goodman}. When coherent, uniform-amplitude light filed is incident on a phase-only spatial light modulator (SLM), it is modulated by a CGH pattern $\phi(x,y)$. This modulated light field then propagates to a screen positioned at a distance $z$ from the SLM, following the propagation function $f_z$. For $f_z$, the band-limited angular spectrum method was used~\cite{Matsushima:09}:
 \begin{align}
    \label{eq:ASM}
    f_z(\phi)&=\iint\mathcal F
     (e^{j\phi(x,y)})H(f_x,f_y)e^{j2\pi(f_xx+f_yy)}df_xdf_y, \\
     \label{eq:H}
     H(f_x,f_y) &= H_{filter} \cdot e^{j 2\pi z \sqrt{\frac{1}{\lambda^2} - (f_x^2 + f_y^2)}}, \\
    H_{filter} &= \begin{cases}
        1 & f_{x}^2+f_{y}^2<\frac{1}{\lambda^2}\\
        0 & \text{otherwise}
    \end{cases},
\end{align} 
where ${\mathcal F}$ is the Fourier transform, $e$ is the Napier number, $j$ is the imaginary unit, $k$ is a wave number, and $f_x$ and $f_y$ are spatial frequencies. Using $f_z$, the amplitude distribution of the reconstructed image $\hat{a}_{target}$ on the screen is denoted as $\hat{a}_{target}=|f_z(\phi)|$. CGH synthesis is the process of determining the optimal $\phi(x,y)$ that minimizes the difference between the reconstructed image $\hat{a}_{target}$ and the desired target image amplitude distribution $a_{target}$.



\subsection{Architecture of Distance-Adaptive CGH Generator}
\begin{figure}
  \centering
  \includegraphics[page=2,width=13cm,pagebox=cropbox,clip]{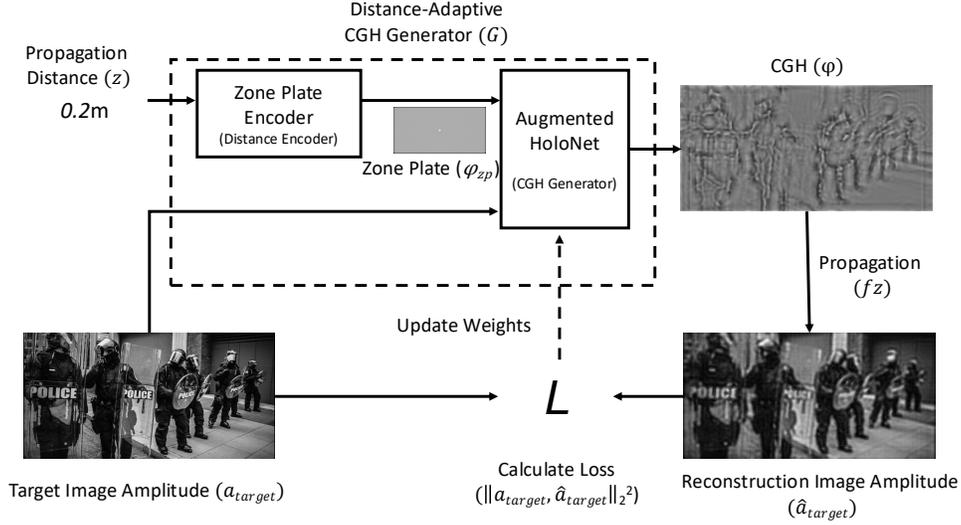}
  \caption{How the CNN learns the weights. When the target image and propagation distance are given, the zone plate encoder first generates a zone plate that serves as an input for the CNN providing a CGH as a single-image output. Then, the accuracy between the image reproduced by diffraction calculations at the given propagation distance and that at the given target image. Based on this result, the weights of the CNN are computed.}
  \label{fig:cnn}
\end{figure}
Our aim is to develop a distance-adaptive CGH generator $G$ (see Fig.~\ref{fig:cnn}), a CNN that inputs a 2D target image $a_{target}$ and a propagation distance $z$, and synthesizes a CGH $\phi(x,y)$ optimized for distance $z$. Using $G$, $\phi(x,y)$ is expressed as $\phi(x,y)=G(a_{target},z)$. Hence, our objective is to solve the following optimization problem:
\begin{equation}
    \begin{aligned}
    \label{eq:objective}
    \underset{G}{min}\mathcal{L}(G) &=\mathcal{L}(a_{target},\hat{a}_{target})\\
                      &=\mathcal{L}(a_{target},|f_{z}(\phi)|)\\
                      &=\mathcal{L}(a_{target},|f_z(G(a_{target},z))|),
\end{aligned}
\end{equation}

where $\mathcal{L}$ is a loss function, and we used the mean-squared error (MSE).
The process of solving this optimization problem can be decomposed into two steps: determining the architecture of $G$ and training $G$ according to Eq.~\eqref{eq:objective}.

Various neural networks can satisfy this objective. One approach is use of conditional generative adversarial networks (cGANs). Gradlow et al.~\cite{gladrow2019digital} and Kang et al.~\cite{kang2021deep} proposed methods for synthesizing CGH using cGANs. Although these methods appear to align with our objective by linking the propagation distance with the latent vector, but they have not effectively demonstrated high-quality 2D image reproduction. Conversely, U-Net based models\cite{citl,yu2022phase} have successfully synthesized CGH with PSNR surpassing 30 dB, yet they cannot directly process scalar propagation distance.

We proposed a U-Net based model capable of managing propagation distance as $G$. To develop such a model, we decided to compose $G$ from two components: a distance encoder and a CGH generator. The distance encoder converts a scalar propagation distance into a 2D image, as U-Net models only accept images as inputs. The CGH generator, an enhanced U-Net model, utilizes the target image and the encoded propagation image to produce CGH.

Various combinations of distance encoders and CGH generators exist. We used a zone plate encoder for distance encoder, and augmented HoloNet for CGH generator because this combination recorded good accuracy empirically. We documented the results of different combinations in the supplemental material.

\subsection{Zone Plate Encoder}
We devised a zone plate encoder to serve as one of the distance encoders for encoding a scalar propagation distance into a 2D image. A zone plate is a well-known CGH for point light sources. Despite its simple shape consisting of several concentric circles, the spacing between the lines changes according to the distance $z$ between the reconstructed image and SLM, making it suitable for representing diffraction phenomena. $\phi_{zp}$ is the output of the zone plate encoder, which is the zone plate corresponding to the propagation distance $z$:
\begin{align}
    \phi_{zp}(x,y) &= 
    \mathrm{arg}(f_z(a_{pl})),
\end{align}
where $a_{pl}$ denotes the amplitude distribution of the point light source. The value of the central point of $a_{pl}$ is 1, and all other regions are 0. $\mathrm{arg}(x)$ represents the angle of the complex number $x$, and $f_z$ is the band-limited angular spectrum method defined in Eq.~\eqref{eq:ASM}.

\subsection{Augmented HoloNet}
\begin{figure}
  \centering
  \includegraphics[page=4,width=13cm,pagebox=cropbox,clip]{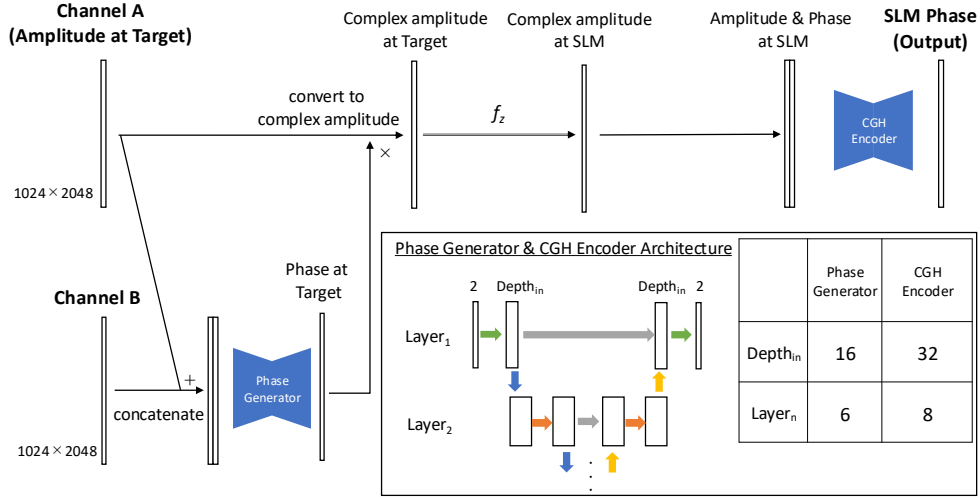}
  \caption{Schematic representation of the augmented HoloNet. Based on the input received in Channels A and B, the phase generator predicts the phase distribution of the target. The complex amplitude distribution on the target plane is calculated using the predicted phase distribution and the amplitude distribution of the target. This is diffracted through $f_z$, as defined by Eq.~\eqref{eq:ASM}, to predict the complex amplitude distribution on the SLM plane, from which the optimal phase distribution on the SLM plane is predicted. Note that $z$ in $f_z$ is the same as that used when calculating Channels A and B. The blue, orange, green, yellow gray arrows represent $4\times4$ convolution and ReLU, $3\times3$ convolution, $1\times1$ convolution, $3\times3$ up-convolution and ReLU, and concatenation, respectively.}
  \label{fig:augholonet}
\end{figure}

The CGH generator must accept two images as input and produce a CGH $\phi$. We propose an augmented HoloNet, which achieves the highest accuracy in our experiments. HoloNet\cite{citl}, a CNN model by Peng et al., generates CGH and, despite having fewer parameters than a 10-layer U-Net, improves PSNR by 10 dB. We modified the target phase generator, one of HoloNet's two U-Nets, to accept two inputs (see Fig.~\ref{fig:augholonet}) and adjusted the propagation distance in the diffraction calculation to match the distance $z$ used for the zone plate. Additionally, we increased the number of layers in both the target phase generator and phase encoder to develop a more versatile model than the original HoloNet.

\subsection{Training the CNN} \label{sec:train_cnn}
In the training process, 3,450 images from the DIV2K\cite{div2k_Agustsson_2017_CVPR_Workshops} and Flickr2K\cite{flickr2k} training datasets were used. For validation, 100 images from the DIV2K validation datasets were used. Although the original HoloNet was trained using only 800 images from DIV2K, we increased the number of training images in this study to capture a greater variety of matching patterns between the images and the propagation distance $z$.

To ease the training process, the propagation distance $z_n$ is randomly selected from a total of $n$ candidates arranged at equal intervals $\beta\lambda$ from 
$z_{0}$ to $z_{max}$, where 
\begin{equation}
\label{eq:z0}
      z_0= d +\alpha \lambda ,
\end{equation}
\begin{equation}
\label{eq:zn}
    z_{n}=z_0+n\beta\lambda .
\end{equation}
The value of $\alpha$ is case-dependent; we chose a value of $\frac{1}{2}$ in this study because it produced good results empirically. The detailed reason for this choice is discussed in Section~\ref{sec:pdis_reason}.  The range of possible values is determined by the value of $\beta$. In this study, we adopt $d=0.2$, $n_{max}=99$, $\beta=2000$.$ (z_{99}\approx0.3$ $ \mathrm{m})$. In the results section, we also describe the changes in accuracy when $d$, $n$, and $\beta$ values are altered. As optical parameters to evaluate numerical reconstruction images, we use $\lambda=520 \mathrm{nm}$, with an SLM pitch size of 6.4 $\mathrm{um}$ and an SLM resolution of $1024\times2048$, consistent with HoloNet implementation. All propagation distances meet Nyquist frequency requirements.

We used MSE as the loss function in this study and employed the Adam optimizer\cite{kingma2017adam}, with consistent values of $\gamma=10^{-4}$, $\beta_{1}=0.99$, $\beta_{2}=0.999$, $\epsilon=10^{-8}$, and $\lambda'=0$ throughout the training process. $\gamma$ is the learning rate, and $\lambda'$ is the weight decay.

In all cases, regardless of the model or input generation method used, 30,000 training iterations were conducted. Validation was performed every 50 iterations, and the PSNR values were recorded. All training and evaluation processes were executed using a Tesla V100 SXM2 graphics processing unit, Python 3.7.13, and Pytorch 1.13.1.
\section{Results}
\subsection{Numerical Results}\label{numerical}
\begin{figure}
  \centering
  \includegraphics[page=32,width=10cm,pagebox=cropbox,clip]{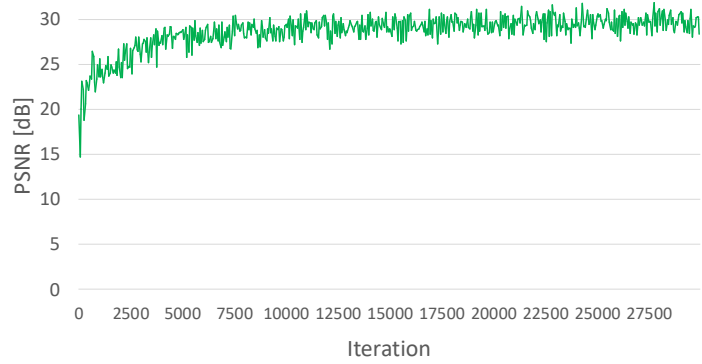}
  \caption{Training results of the four proposed patterns. Our model was trained for 30,000 iterations, and the PSNR values evaluated by the validation dataset were recorded on the vertical axis every 50 iterations in training. }
  \label{fig:val_chart}
\end{figure}

\begin{figure}
  \centering
  \includegraphics[page=33,width=10cm,pagebox=cropbox,clip]{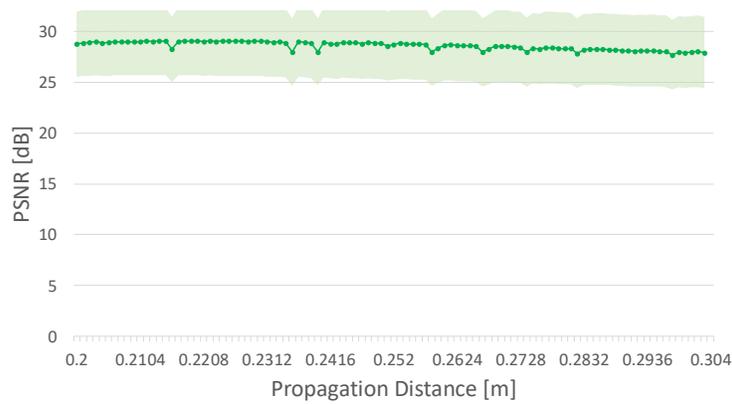}
  \caption{Evaluated accuracy of the CGH output by our model for each propagation distance. The propagation distances are the 100 values defined  in section~\ref{sec:train_cnn}. The accuracy of the model at each propagation distance was calculated using the mean of 100 validation images. The background colors represent the standard deviations}
  \label{fig:sanpuzu}
\end{figure}

\begin{figure}
  \centering
  \includegraphics[width=13cm]{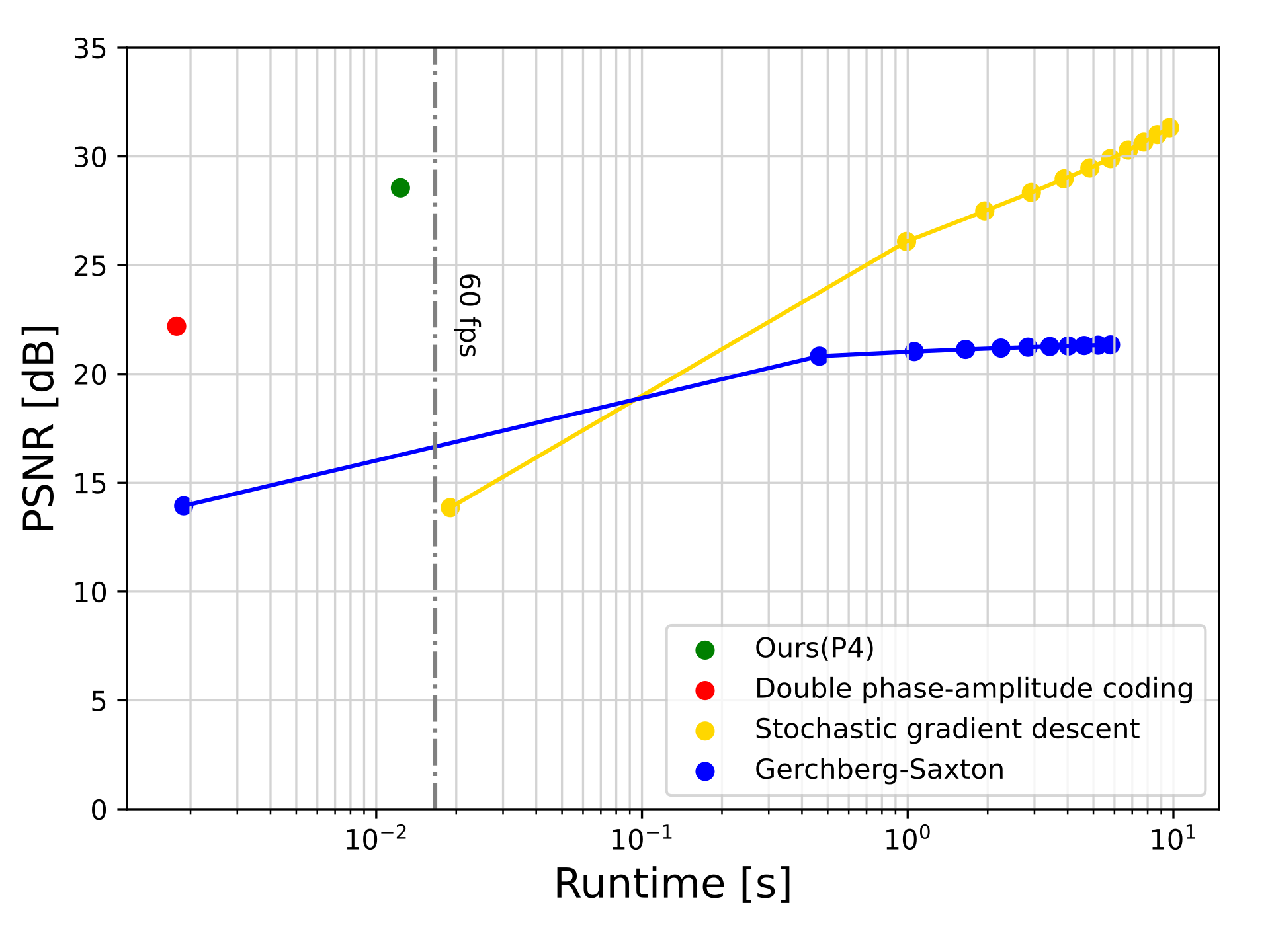}
  \caption{Comparison of the CGH generation runtime and accuracy between the proposed and existing methods. The values for each point use the average of the conditions when the CGH is generated using a total of 10,000 sets consisting of 100 distance patterns and 100 validation images.}
  \label{fig:other_model}
\end{figure}

\begin{figure}
  \centering
  \includegraphics[page=35,width=13cm,pagebox=cropbox,clip]{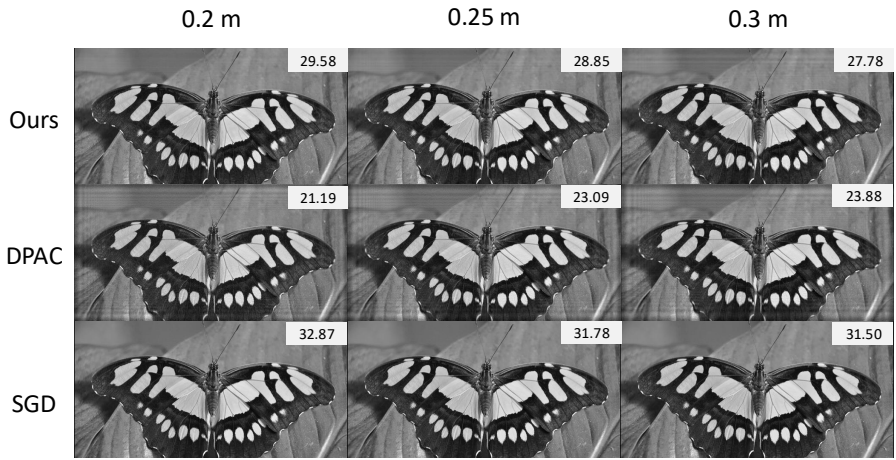}
  \caption{Results of the reconstruction simulation using the output CGH, obtained by inputting images from the validation dataset into each model pattern after 30,000 iterations of training.}
  \label{fig:final_output}
\end{figure}

We trained and evaluated a distance-adaptive CGH generator, with results depicted in Fig.~\ref{fig:val_chart}. The average PSNR value over iterations was 28.63 dB. Our model demonstrated stable accuracy across all trained target distances (Fig.~\ref{fig:sanpuzu}), without showing higher accuracy for specific distances, indicating no overfitting. The average PSNR value over 100 distances was 28.64 dB.

The results comparing the CGH generation runtime and accuracy of our method with other CGH generation methods are illustrated in Fig.~\ref{fig:other_model}. Our method exhibited high accuracy relative to existing methods. Additionally, generation speeds exceeding 60 fps were achieved, indicating its capability for high-speed generation. The PSNR and runtime for each point were averaged over 10,000 combinations, comprising 100 distance patterns and 100 validation images. Numerical reconstruction images for each model are shown in Fig.~\ref{fig:final_output}. Notably, in our implementation, Eq. \eqref{eq:H} is pre-calculated for the number of distance patterns beforehand, but this pre-calculation time is excluded from the runtime, as discussed in the Discussion section.
\subsection{Experimental Results}
\begin{figure}[H]
  \centering
  \includegraphics[page=36,width=13cm,pagebox=cropbox,clip]{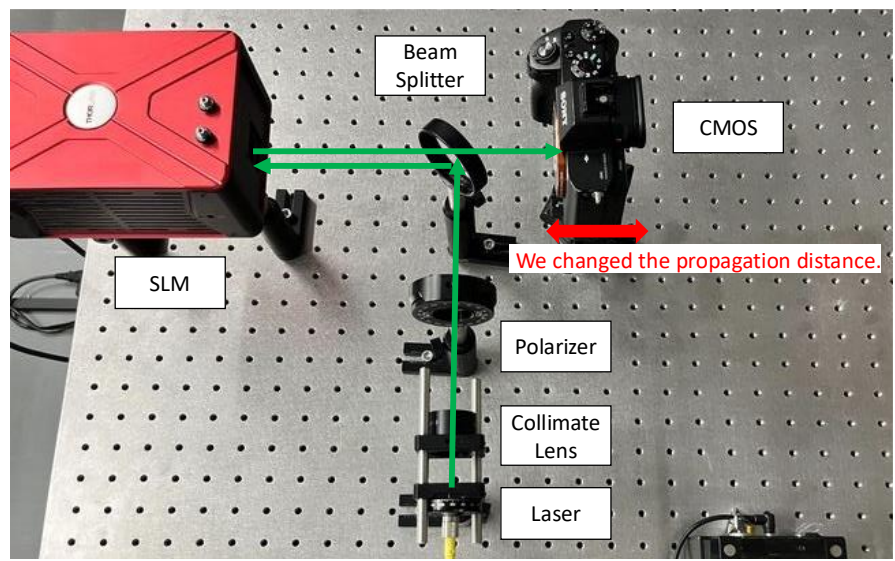}
  \caption{Optical setup used in the experiment: Light emitted from a laser source is collimated by a lens, modified, and incident on the SLM. The light modulated by the SLM travels a propagation distance $z$ to reach the CMOS sensor.}
  \label{fig:setup}
\end{figure}

We assessed our model's accuracy under physical conditions through a CGH reconstruction experiment using a Thorlabs-EXULUS-4K1 SLM (3.74 µm pitch size) and a 517.9 nm wavelength laser. The model was tuned accordingly, and the CGH resolution was set to 2048 × 1024, projected onto the SLM's central part. A SONY $\alpha$7II camera captured the reconstructed images. The optical setup is shown in Fig.~\ref{fig:setup}. To alter the propagation distance, we repositioned the camera.

\begin{figure}[H]
  \centering
  \includegraphics[page=37,width=13cm,pagebox=cropbox,clip]{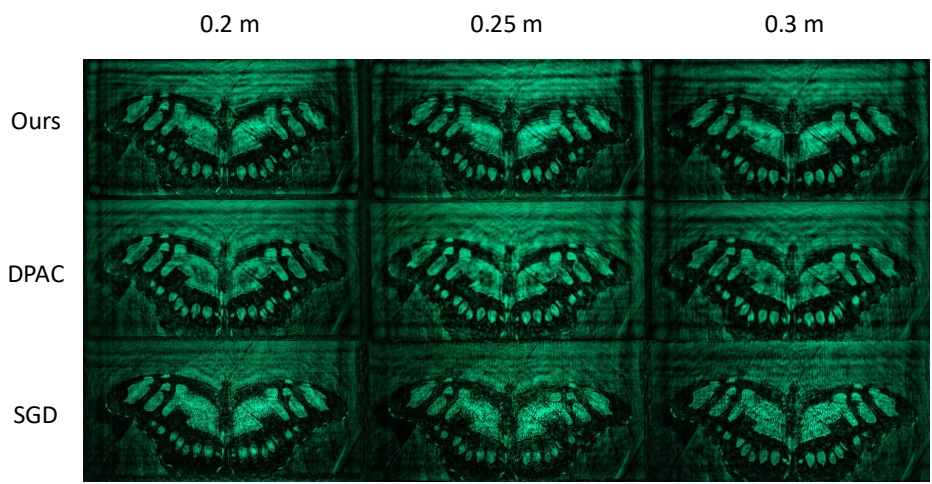}
  \caption{Optical CGH reconstruction generated using Ours, DPAC, and SGD methods at propagation distances of 0.2, 0.25, and 0.3 m.}
  \label{fig:optical_result}
\end{figure}

Fig.~\ref{fig:optical_result} displays optical reconstruction images at propagation distances of 0.2, 0.25, and 0.3~m using DPAC, SGD, and our method. Our reconstructed images exhibited less blurriness compared to DPAC and higher contrast and clarity than SGD. The accuracy of the reconstructed images remained consistent despite varying propagation distances, indicating that our model effectively met the objective.

\subsection{Model Versatility}
\begin{figure}
  \centering  \includegraphics[width=13cm]{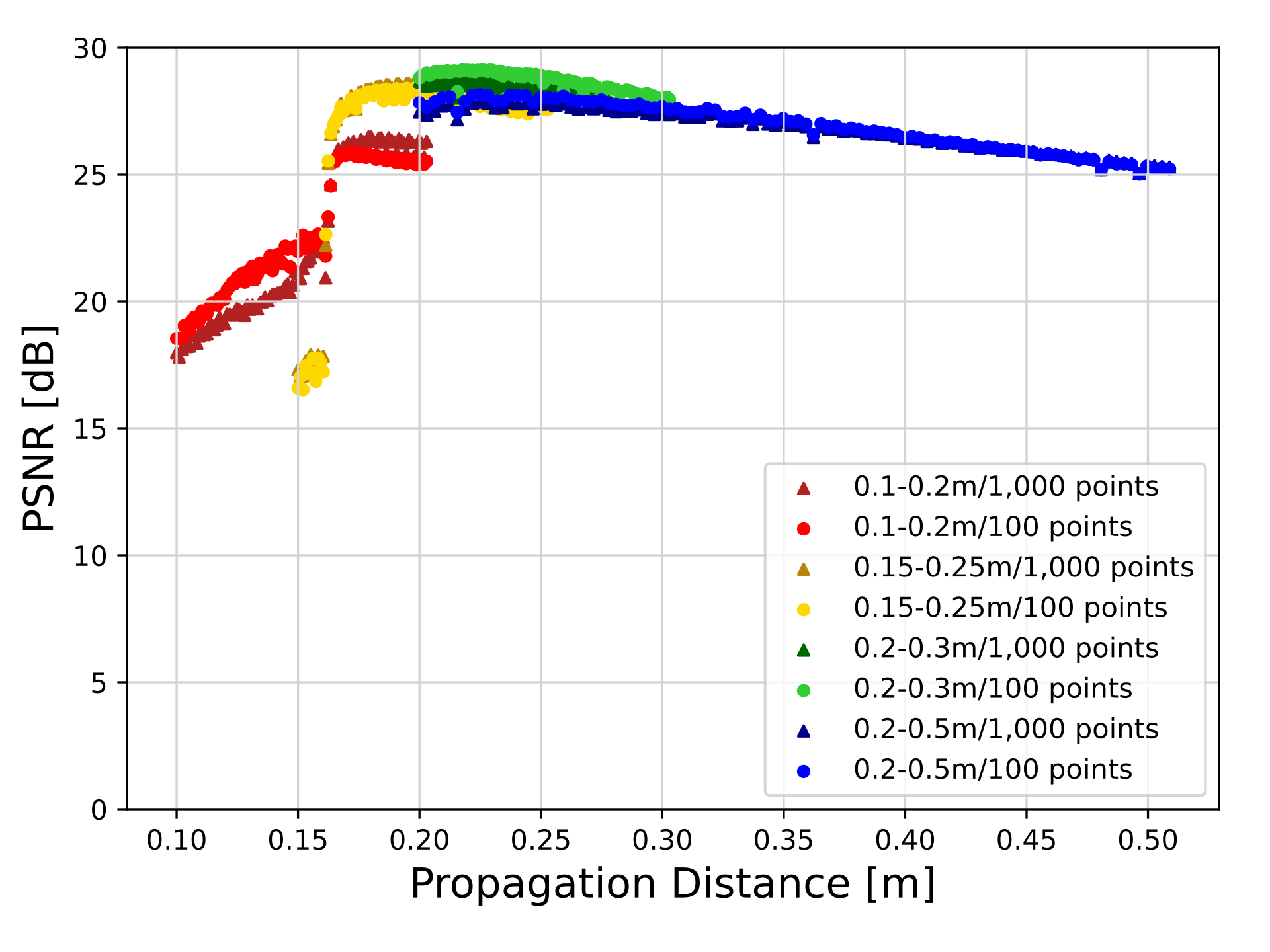}
  \caption{Comparison of PSNR between patterns with varying numbers of training distances and varying as starting and ending positions. All patterns employed learning data after 30,000 iterations. Evaluations were performed solely on the distances included within 100 learning points, even when there were 1,000 learning distances.}
  \label{fig:change_dis}
\end{figure}

To investigate the limitation of this model, the propagation distance to be trained was varied, and the accuracy of CGH generation was evaluated. The results are shown in Fig.~\ref{fig:change_dis}. Focusing on points where the propagation distance is 0.2 m or more, we observed that when the number of propagation distances to be learned increased tenfold, the accuracy dropped by an average of 0.3 dB. Moreover, extending the propagation distance decreased the accuracy. However, in all instances, the PSNR recorded was above 25 dB, suggesting that our proposed method remains better even when compared with techniques such as DPAC. The accuracy drops sharply at points less than 0.2 m; however, it has been confirmed that the same phenomenon occurs at this distance with both DPAC and HoloNet\cite{citl}.

\begin{figure}
  \centering
  \includegraphics[page=9,width=13cm,pagebox=cropbox,clip]{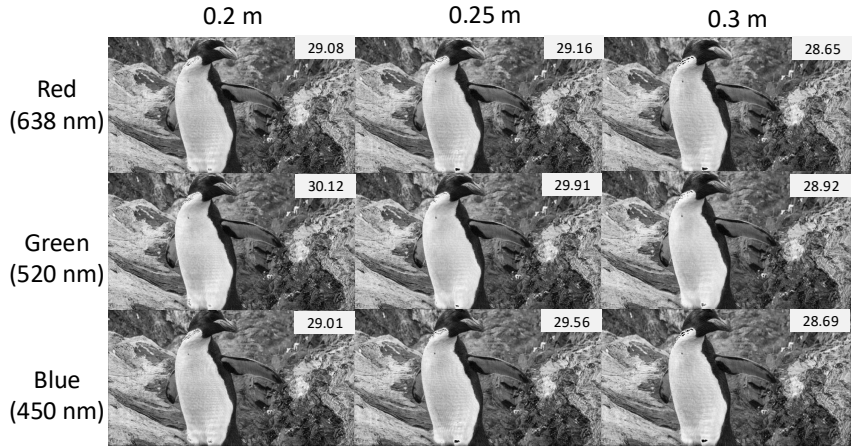}
  \caption{Comparison of CGH reconstruction images for each wavelength and distance. After completing 30,000 iterations of training for each wavelength, the CGH was generated by providing the target image and propagation distance.}
  \label{fig:colors}
\end{figure}

To verify the model's versatility, we tested its accuracy using green (520 nm), red (638 nm), and blue (450 nm) wavelengths. We validated 10,000 pairs of images and propagation distances. The average PSNR values for red, green, and blue were 27.20, 28.64, and 27.52 dB, respectively. Reconstructed images for each color and propagation distance are shown in Fig.~\ref{fig:colors}. These results indicate the model's effectiveness across multiple wavelengths.

\section{Discussion}
\label{sec:discuss}
\subsection{Performance of the Proposed Model}\label{sec.modelpf}
As described in the Results section, our model can maintain the PSNR at an average of 28.64 dB even when the propagation distance changes. This is considered a sufficient result, as it is comparable to that of other CNNs developed for generating CGH. However, only four combinations of distance encoder and CGH generator were tested in this study. It is possible that the accuracy could be further improved using other models. Furthermore, a study has emerged that significantly improves the precision of HoloNet \cite{Dong:23}, suggesting that by applying these findings, it would be possible to enhance accuracy further.

The decline in accuracy below 0.2 m is problematic for this model, despite its universality in propagation distance. Although the stochastic gradient descent (SGD) method does not exhibit this issue, it has been observed in the HoloNet model, indicating that suitable optimization could maintain accuracy within this range. However, accomplishing this with the HoloNet model is challenging, warranting future research to develop a solution.

We trained and evaluated by varying the propagation distance from 0.2~m to 0.5~m, with a maximum range of 0.3~m. In practical CGH applications, the propagation distance variation may exceed 0.3~m, making performance unpredictable. CGH accuracy is generally influenced by all factors: light wavelength, SLM pitch size, and propagation distance. Accuracy could vary significantly with different parameter substitutions.

\subsection{Effects and Solution for Misalignment Between Simulation Environments and Psychical Systems}
Our model was trained in an ideal simulation environment, assuming constant light amplitude on every SLM pixel, no aberrations in the optical system, and lossless light propagation. Real-world optical systems inevitably deviate from these ideal conditions, which we believe causes the lower quality of reconstructed images in physical experiments compared to simulations for all models, including ours, DPAC, and SGD. Despite these discrepancies, the optical reconstructed images remained recognizable, suggesting that models trained under ideal conditions can tolerate some error. To address this, camera-in-the-loop training, as demonstrated by Peng et al.~\cite{citl}, can be utilized. However, we did not implement this in our study, as we believe the goal of generating distance-adaptive CGH has been adequately achieved.

\begin{figure}
  \centering
  \includegraphics[width=13cm]{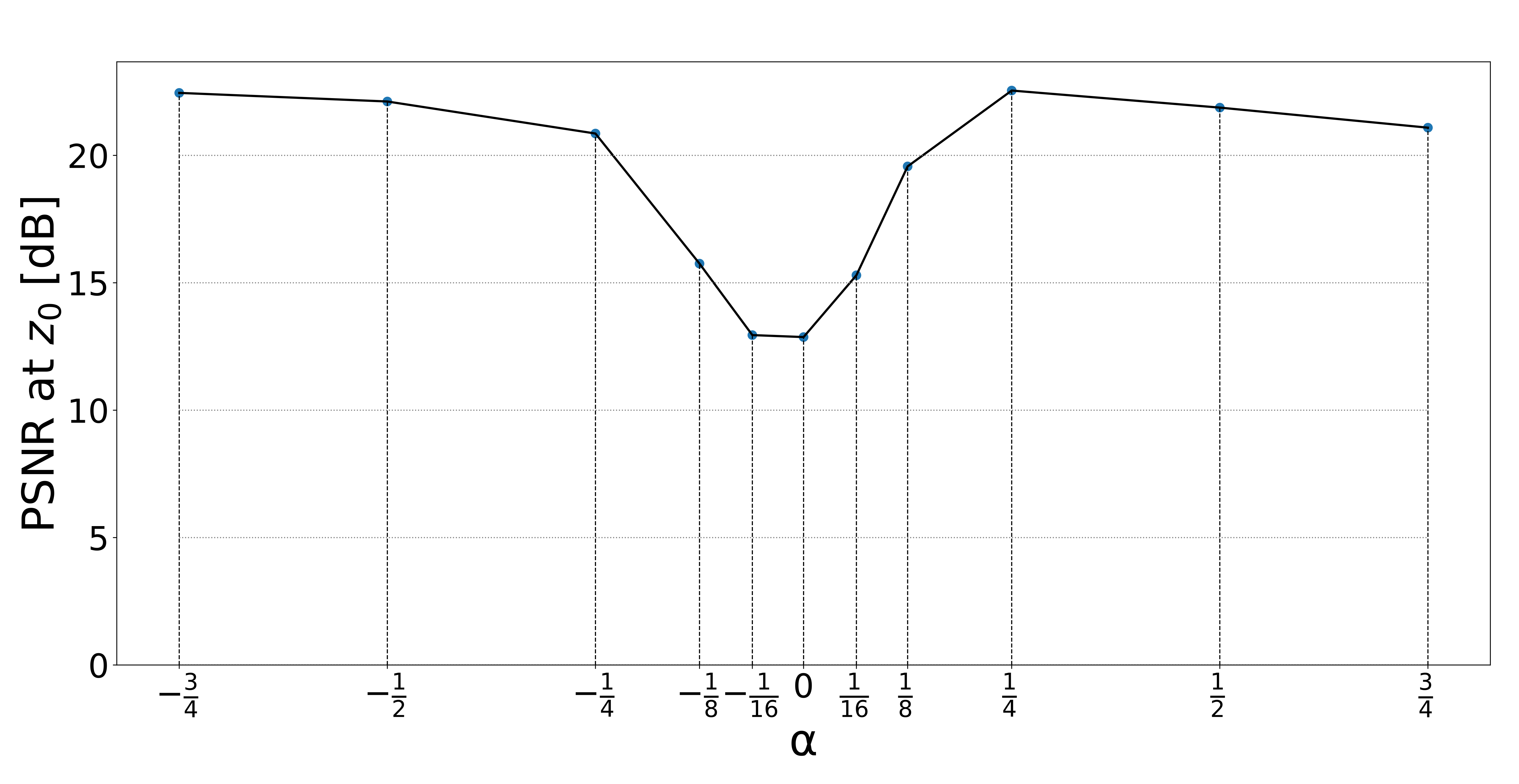}
  \caption{Accuracy of our proposed method (P4) when the value of $\alpha$ is changed. Eleven patterns of $\alpha$ values, $-\frac{3}{4}$, $-\frac{1}{2}$, $-\frac{1}{4}$, $-\frac{1}{8}$ ,$-\frac{1}{16}$, $0$, $\frac{1}{16}$, $\frac{1}{8}$, $\frac{1}{4}$, $\frac{1}{2}$, and $\frac{3}{4}$, were adopted as training targets. The values of $\beta$, $d$, and $n$ of Eq.~\eqref{eq:zn} were 200,000, 0.2, and 99 for each alpha, resulting in a total of 1,100 points used as training propagation distances. Although all 1,100 points are targeted for both learning and evaluation, the points presented in this figure are only $z_0$ for each $\alpha$.}
  \label{fig:alpha}
\end{figure}

\subsection{Purpose of Choosing the Propagation Distance}\label{sec:pdis_reason}
As shown in Section~\ref{sec:train_cnn}, the propagation distance used in this study was randomly selected from regular interval discretization to ease the training process. We found that there are propagation distances $z'$ at which the PSNR decreases significantly. Even when the same discretization step $\beta$ is used, the performance varies depending on the initial starting point, $z_0$ (i.e.,~$\alpha$ defined at Eq.\eqref{eq:z0}). For example, Fig.~\ref{fig:alpha} describes our model accuracy when it was trained with  1,100 points defined by changing the $\alpha$ values in eleven patterns ($\alpha=$ $-\frac{3}{4}$, $-\frac{1}{2}$, $-\frac{1}{4}$, $-\frac{1}{8}$,$-\frac{1}{16}$, $0$, $\frac{1}{16}$, $\frac{1}{8}$, $\frac{1}{4}$, $\frac{1}{2}$, $\frac{3}{4}$). This indicates that the PSNR decreases significantly only when $\alpha$ approaches $0$ ($z'$ mod $\lambda =0 $). In this case, it is preferable to use $\alpha=\frac{1}{4}$, $\frac{1}{2}$ or $\frac{3}{4}$. In our experiments, we found that $\alpha$ is influenced by the pitch size of the SLM, and particular care must be taken when changing it. Owing to this limitation, the discretization step cannot be less than $1\lambda$ order to avoid low PSNR steps. For holographic projectors, applications do not require  < $1\lambda$ distance adjustments. However, it is worth noting that distance selection is an important process in CNN training.

\subsection{Caching of Band-Limited ASM Kernels}
As mentioned in Section~\ref{numerical}, this study cached the diffraction calculation kernel corresponding to the propagation distance to be trained, which corresponds to Eq.~\eqref{eq:H}. This is considered an appropriate implementation, assuming that in Peng et al.~\cite{citl}, where the propagation distance is fixed, the kernel is calculated once and then reused in subsequent calculations.
Using this method, as the number of propagation distances to be learned increases, more time is required for pre-computation. If there are 100 distance patterns, this takes approximately one minute. However, in actual products, it is unlikely that the distances that need to be accommodated would change frequently; therefore, the frequency of performing this pre-computation is presumed to be very low. Hence, the execution time of this pre-computation should not pose a significant problem.

\subsection{Anticipated Application}
As mentioned in the introduction, potential applications of this study include holographic projectors\cite{Makowski:12,Buckley:11,Shimobaba:13} and holographic near-eye displays\cite{MS_near-eye_holo, Kim2022, near-eye-aberration}. Unlike conventional projectors, the positions of the screen and holographic projector may change frequently depending on the usage scenario. However, it is challenging to adapt to such situations using CGH generation methods that do not allow the propagation distance to be specified.

In our study, as demonstrated in the Results section, such issues did not arise within the tested range of distances. Holographic near-eye displays are expected to be worn by many people. In this case, it would be desirable to adjust the position of the reconstructed image appropriately according to the wearer's facial shape and the content being displayed. As previously mentioned, our study enables changing the propagation distance at $1\lambda$ intervals, which we believe satisfies this requirement. 

Moreover, a highly accurate and real-time CGH generation method that can specify the propagation distance is also desired in the field of acoustic holography. The versatile CGH generation using the CNN employed in this study may contribute to this field\cite{gs-pat,diff-pat}. 

\section{Conclusion}
In this study, we developed a CNN that outputs high-precision CGHs despite specifying the propagation distance as an input. Although some models successfully added CNN models to add conditions, we believe that this is an important study that directly converts optical parameters to images and validates the range of effective conditions. This suggests the possibility of freely modifying other parameters using similar methods. A future challenge is to develop a general-purpose CNN model that can adapt to changes in other optical calculation parameters, such as wavelength and pitch size, as well as to human eye aberrations.

\begin{backmatter}
\bmsection{Funding}
This work was funded by Pixie Dust Technologies, Inc.

\bmsection{Acknowledgments}
We would like to thank Editage (www.editage.jp) for English language editing.

\bmsection{Disclosures}
The authors declare no potential conflicts of interest.
\bmsection{Data Availability}
Data underlying the results presented in this paper will be made accessible on Zenodo [XXX] and Github [XXX]
\end{backmatter}


\bibliography{sample}

\begin{thebibliography}{10}
\newcommand{\enquote}[1]{``#1''}

\bibitem{cgh_review}
E.~Sahin, E.~Stoykova, J.~M\"{a}kinen, and A.~Gotchev, \enquote{Computer-generated holograms for {3D} imaging: a survey,} {\protect\JournalTitle{ACM Comput. Surv.}} \textbf{53}, 1--35 (2020).

\bibitem{MS_near-eye_holo}
A.~Maimone, A.~Georgiou, and J.~S. Kollin, \enquote{Holographic near-eye displays for virtual and augmented reality,} {\protect\JournalTitle{ACM Trans. Graph.}} \textbf{36}, 8501--8506 (2017).

\bibitem{Kim2022}
J.~Kim, M.~Gopakumar, S.~Choi, Y.~Peng, W.~Lopes, and G.~Wetzstein, \enquote{Holographic glasses for virtual reality,} in \emph{ACM SIGGRAPH 2022 Conference Proceedings,}  (Association for Computing Machinery, New York, NY, USA, 2022), SIGGRAPH '22, pp. 1--9.

\bibitem{near-eye-aberration}
K.~Yamamoto, I.~Suzuki, K.~Namikawa, K.~Sato, and Y.~Ochiai, \enquote{Interactive eye aberration correction for holographic near-eye display,} in \emph{Proceedings of the Augmented Humans International Conference 2021,}  (Association for Computing Machinery, New York, NY, USA, 2021), AHs '21, p. 204–214.

\bibitem{Makowski:12}
M.~Makowski, I.~Ducin, K.~Kakarenko, J.~Suszek, M.~Sypek, and A.~Kolodziejczyk, \enquote{Simple holographic projection in color,} {\protect\JournalTitle{Opt. Express}} \textbf{20}, 25130--25136 (2012).

\bibitem{Buckley:11}
E.~Buckley, \enquote{Holographic laser projection,} {\protect\JournalTitle{J. Display Technol.}} \textbf{7}, 135--140 (2011).

\bibitem{Shimobaba:13}
T.~Shimobaba, M.~Makowski, T.~Kakue, M.~Oikawa, N.~Okada, Y.~Endo, R.~Hirayama, and T.~Ito, \enquote{Lensless zoomable holographic projection using scaled fresnel diffraction,} {\protect\JournalTitle{Opt. Express}} \textbf{21}, 25285--25290 (2013).

\bibitem{gabor}
D.~GABOR, \enquote{A new microscopic principle,} {\protect\JournalTitle{Nature (London)}} \textbf{161}, 777--778 (1948).

\bibitem{GS}
G.~R. W., \enquote{A practical algorithm for the determination of the phase from image and diffraction plane pictures,} {\protect\JournalTitle{Optik}} \textbf{35}, 237--246 (1972).

\bibitem{Zhang:17}
J.~Zhang, N.~P\'{e}gard, J.~Zhong, H.~Adesnik, and L.~Waller, \enquote{{3D} computer-generated holography by non-convex optimization,} {\protect\JournalTitle{Optica}} \textbf{4}, 1306--1313 (2017).

\bibitem{chakravarthula2019wirtinger}
P.~Chakravarthula, Y.~Peng, J.~Kollin, H.~Fuchs, and F.~Heide, \enquote{Wirtinger holography for near-eye displays,} {\protect\JournalTitle{ACM Trans. Graph.}} \textbf{38}, 213 (2019).

\bibitem{citl}
Y.~Peng, S.~Choi, N.~Padmanaban, and G.~Wetzstein, \enquote{Neural holography with camera-in-the-loop training,} {\protect\JournalTitle{ACM Trans. Graph.}} \textbf{39}, 1--14 (2020).

\bibitem{shimobaba2022deep}
T.~Shimobaba, D.~Blinder, T.~Birnbaum, I.~Hoshi, H.~Shiomi, P.~Schelkens, and T.~Ito, \enquote{Deep-learning computational holography: A review,} {\protect\JournalTitle{Frontiers in Photonics}} \textbf{3}, 8 (2022).

\bibitem{Horisaki:18}
R.~Horisaki, R.~Takagi, and J.~Tanida, \enquote{Deep-learning-generated holography,} {\protect\JournalTitle{Appl. Opt.}} \textbf{57}, 3859--3863 (2018).

\bibitem{u_net}
O.~Ronneberger, P.~Fischer, and T.~Brox, \enquote{{U-Net}: Convolutional networks for biomedical image segmentation,} in \emph{Medical Image Computing and Computer-Assisted Intervention -- MICCAI 2015,}  N.~Navab, J.~Hornegger, W.~M. Wells, and A.~F. Frangi, eds. (Springer International Publishing, Cham, 2015), pp. 234--241.

\bibitem{Wang:19}
K.~Wang, J.~Dou, Q.~Kemao, J.~Di, and J.~Zhao, \enquote{{Y-Net}: a one-to-two deep learning framework for digital holographic reconstruction,} {\protect\JournalTitle{Opt. Lett.}} \textbf{44}, 4765--4768 (2019).

\bibitem{Dong:23}
Z.~Dong, C.~Xu, Y.~Ling, Y.~Li, and Y.~Su, \enquote{Fourier-inspired neural module for real-time and high-fidelity computer-generated holography,} {\protect\JournalTitle{Opt. Lett.}} \textbf{48}, 759--762 (2023).

\bibitem{yu2022phase}
T.~Yu, S.~Zhang, W.~Chen, J.~Liu, X.~Zhang, and Z.~Tian, \enquote{Phase dual-resolution networks for a computer-generated hologram,} {\protect\JournalTitle{Optics Express}} \textbf{30}, 2378--2389 (2022).

\bibitem{shi2021towards}
L.~Shi, B.~Li, C.~Kim, P.~Kellnhofer, and W.~Matusik, \enquote{Towards real-time photorealistic 3d holography with deep neural networks,} {\protect\JournalTitle{Nature}} \textbf{591}, 234--239 (2021).

\bibitem{shi2022end}
L.~Shi, B.~Li, and W.~Matusik, \enquote{End-to-end learning of 3d phase-only holograms for holographic display,} {\protect\JournalTitle{Light: Science \& Applications}} \textbf{11}, 247 (2022).

\bibitem{Shui:22}
X.~Shui, H.~Zheng, X.~Xia, F.~Yang, W.~Wang, and Y.~Yu, \enquote{Diffraction model-informed neural network for unsupervised layer-based computer-generated holography,} {\protect\JournalTitle{Opt. Express}} \textbf{30}, 44814--44826 (2022).

\bibitem{goodman}
J.~W. Goodman, \emph{Introduction to Fourier optics} (Roberts, Englewood, 2005), 3rd ed.

\bibitem{Matsushima:09}
K.~Matsushima and T.~Shimobaba, \enquote{Band-limited angular spectrum method for numerical simulation of free-space propagation in far and near fields,} {\protect\JournalTitle{Opt. Express}} \textbf{17}, 19662--19673 (2009).

\bibitem{gladrow2019digital}
J.~Gladrow, \enquote{Digital phase-only holography using deep conditional generative models,} {\protect\JournalTitle{arXiv preprint arXiv:1911.00904}}  (2019).

\bibitem{kang2021deep}
J.-W. Kang, B.-S. Park, J.-K. Kim, D.-W. Kim, and Y.-H. Seo, \enquote{Deep-learning-based hologram generation using a generative model,} {\protect\JournalTitle{Applied Optics}} \textbf{60}, 7391--7399 (2021).

\bibitem{div2k_Agustsson_2017_CVPR_Workshops}
E.~Agustsson and R.~Timofte, \enquote{Ntire 2017 challenge on single image super-resolution: Dataset and study,} in \emph{The IEEE Conference on Computer Vision and Pattern Recognition (CVPR) Workshops,}  (2017).

\bibitem{flickr2k}
\enquote{Flickr2k dataset,} [Online]. Available: \url{http://cv.snu.ac.kr/research/EDSR/Flickr2K.tar}.

\bibitem{kingma2017adam}
D.~P. Kingma and J.~Ba, \enquote{Adam: A method for stochastic optimization,} ArXiv:1412.6980 (2017).

\bibitem{gs-pat}
D.~M. Plasencia, R.~Hirayama, R.~Montano-Murillo, and S.~Subramanian, \enquote{{GS-PAT}: high-speed multi-point sound-fields for phased arrays of transducers,} {\protect\JournalTitle{ACM Trans. Graph.}} \textbf{39}, 1--12 (2020).

\bibitem{diff-pat}
T.~Fushimi, K.~Yamamoto, and Y.~Ochiai, \enquote{Acoustic hologram optimisation using automatic differentiation,} {\protect\JournalTitle{Scientific reports}} \textbf{11}, 12678--12678 (2021).

\end{thebibliography}






\end{document}